# Implications of Zoning Ordinances for Rural Utility-Scale Solar Deployment and Power System Decarbonization in the Great Lakes Region

Papa Yaw Owusu-Obeng [a,*], Sarah Banas Mills [b,c], Michael T. Craig [a,d]

[a] School for Environment and Sustainability, University of Michigan, Ann Arbor, MI 48109, USA

[b] Graham Sustainability Institute, University of Michigan, Ann Arbor, MI 48104

[c] A. Alfred Taubman College of Architecture and Urban Planning, University of Michigan, Ann Arbor, MI 48109

[d] Department of Industrial and Operations Engineering, University of Michigan, Ann Arbor, MI 48109, USA

***Corresponding author:** Papa Yaw Owusu-Obeng

**Email:** owusup@umich.edu

## Abstract

Local zoning ordinances across the United States can restrict development of energy infrastructure, including utility-scale solar photovoltaics (PV). While ordinances may be developed for legitimate purposes to protect public health and safety, they could impede or increase costs of power sector decarbonization. We quantify the role of utility-scale solar zoning ordinances on power sector decarbonization across the Great Lakes region (Illinois, Indiana, Michigan, Minnesota, Ohio, and Wisconsin) by integrating 2,474 unique rural zoning ordinances (covering over 90% of county subdivisions in each state) into a power system planning model. Our analysis focuses on deploying utility-scale PV on agricultural lands, where most existing solar installations in our region are located. Relative to a hypothetical counterfactual without zoning ordinances, existing zoning ordinances cause an 18% decline in utility-scale PV investment (8 GW) and cost ($4.8 billion), primarily due to "silent" ordinances that implicitly block solar development by failing to address its land-use. Investment shifts from PV to energy storage (900 MW), wind (300 MW), and natural gas plants (3 GW). Starker declines in PV investment occur at the state level, with Michigan and Wisconsin seeing a 42% reduction. Our results underscore the need for planning that aligns local zoning laws with state and regional goals.

## Introduction

Power system planners and utilities must rapidly pursue decarbonization plans to meet net zero emissions targets and mitigate climate change. Scaling up investment in renewable generators, including solar power, is crucial for achieving net zero emission pathways [1]. In 2021, utility-scale PV was the largest source of capacity added to the US, with an annual growth rate of 25% compared to the previous year [2]. Solar PV costs have rapidly declined, with more cost reductions expected in the near future [3]. Besides cost-effectiveness, solar power can provide diverse benefits, including grid support and ancillary services [4], scalability [5], job creation [6], and shorter construction times [7]. Given these advantages and the declining costs, solar power is expected to continue to grow in the near- and long-term. Based on the contracted capacity additions, the US utility-scale PV market is projected to add 437 GW of utility-scale PV between 2022 and 2032 [8]. For the purposes of this analysis, utility-scale solar is defined as PV installations typically sized at 5 megawatts (MW)[2].

Amidst these benefits, siting increasing quantities of utility-scale PV facilities faces several challenges, including unsuitability of land types, competing land uses, buffer requirements around solar and other infrastructures, and potential environmental impacts [9,10]. In this paper, we focus on a locally-driven challenge to siting utility-scale PV: land-use regulations embedded within zoning ordinances [11–14]. Zoning ordinances are the primary legal instruments through which local governments regulate land use within their jurisdictions, which may include counties, townships, municipalities, or villages. State-level policy determines which level(s) of government have zoning authority[15]. Typically, county-level zoning ordinances apply to all unincorporated rural areas, but especially in New England and the Midwest, unincorporated towns or townships, rather than counties, may zone. In many states, some local governments have chosen to remain unzoned, and state-level policies determine whether they may apply other land development ordinances to regulate renewable energy projects[16].

Within zoning ordinances, land use for utility-scale solar can be expressly permitted, prohibited, or not mentioned at all. If solar energy systems are mentioned, the ordinance may include specific provisions that either allow for the development of utility-scale solar facilities, potentially with certain conditions, or explicitly ban them. Bans on utility-scale solar when explicitly mentioned are often specific to zoning districts, such as agricultural areas where preserving farmland or maintaining rural character is prioritized[17]. An assessment of 32 states, for instance, identified 7 to have banned solar installations on farmlands[18].

When ordinances do not mention land use for utility-scale solar, they are referred to as “silent” on solar. Such “silent” ordinances on solar effectively hinder utility-scale solar development in several ways. First, in just more than half of the states[19], zoning laws typically prohibit land uses unless they are explicitly allowed[20]. Consequently, in these states developers cannot build utility-scale solar projects in areas where the use is not expressly permitted. Second, while this silence could be addressed by enacting a new ordinance, that process can significantly increase project development costs[21], lead to unpredictable approval processes, strong opposition, and significant delays. A study by Nolon highlights significant increases in project development costs when zoning ordinances lack clarity or fail to define utility-scale PV[21]. Prominent examples include the proposed 800 MW Blackwater Solar Project in Sussex County and the 165 MW Painter Creek Solar Project in Greenville County, both in Virginia [22,23].

When ordinances clearly delineate solar, typical provisions set clear guidelines for the size and layout of utility-scale PV facilities, such as setbacks from roads, participating property lines, non-participating property lines, lot size minimums and maximums and lot area coverage. These provisions minimize conflicts with other land uses and address concerns related to visual impacts, noise, land conservation, and public health and safety [24]. Well-defined zoning ordinances streamline permitting processes and reduce uncertainties for developers, facilitating the development of new solar projects[25]. It is important to clarify that while the term "ordinance" refers to the entire body of regulations enacted by a local government, individual rules within an ordinance are more accurately termed "provisions" or "regulations." However, in the literature, there is some confusion, with researchers like Lopez et al. sometimes referring to these zoning provisions as "ordinances" themselves[26].

Several studies agree that zoning regulations limit the availability of land for meeting clean energy targets [9,21,27,28]. Recent trends in renewable energy zoning show a rise in restrictive ordinances for utility-scale PV originating from ordinances silent on solar and total bans on deployment [29]. Few studies have been conducted to quantify the specific impacts of local zoning ordinances, particularly those silent on solar, on the cost of power system decarbonization. A study by Lopez et al. [27] highlights that setback requirements represent a significant obstacle to wind deployment, leading to a 20% reduction in potential deployment across the United States. Another study [28] demonstrates that strict land-use regulations can impede transmission access, limit deployable capacity, and increase the cost of electricity generated by wind systems, especially in high electrification scenarios. Lopez et al. [26] further emphasize the potential of zoning ordinances to reduce solar resources by up to 38%.

However, these studies survey only a limited number of zoning jurisdictions or ordinances, and then replicate the solar provisions in these ordinances across multiple jurisdictions. For instance, Lopez et al.[30] survey 295 zoning jurisdictions for solar on a national level, whereas the Great Lakes region alone includes over 2,474 unique zoning jurisdictions [29]. Additionally, these studies do not translate zoning ordinances to impacts on deployment decisions, but instead on available sites for deployment [26–28,31,32]. Several assessments conjecture that given the vast size of solar resource, land availability does not constrain PV deployment [31–33]. But these studies do not consider zoning ordinances, which could impose substantial limits on land available for development on a local and regional level.

In this study, we examine the effect of local zoning ordinances on power sector decarbonization through a focus on utility-scale PV deployment in rural communities in the Great Lakes region of Illinois, Indiana, Michigan, Minnesota, Ohio, and Wisconsin. We specifically consider PV deployment on agricultural lands, which have emerged as significant sites for utility-scale solar deployment due to their flat terrain, availability, and proximity to existing grid infrastructure [34] **(SI Figure S5)**. We first create a zoning database of unprecedented depth by collecting and characterizing the status of solar zoning for 9,997 county subdivisions, more than 90% of county subdivisions across our study region. This database includes 2,474 unique zoning jurisdictions, 8 times more jurisdictions for a six-state region than prior national-level research by Lopez et al [26]. Each zoning jurisdiction has a unique zoning ordinance with several provisions regulating solar. We then couple local zoning ordinances with local solar resource and transmission interconnection costs to generate supply curves of potential solar development sites. These sites are limited to agricultural lands in rural areas, mirroring existing deployment trends in our study region [35]. We feed these supply curves into a power system planning, or capacity expansion (CE), model. The CE model minimizes total annualized system costs across our Great Lakes region by optimizing generator and transmission additions. Utility-scale PV additions in the CE model are constrained by our generated supply curves. To evaluate the impact of zoning ordinances on utility-scale solar deployment for decarbonization, we run our CE model for three zoning policy scenarios: a scenario that ignores zoning ordinances in siting decisions (ignoring zoning"); a scenario that enforces existing zoning ordinances ("current zoning zoning"), and a scenario that extends existing ordinances to jurisdictions that currently lack zoning ordinances ("expanded zoning"). The ignoring zoning scenario is a hypothetical baseline or counterfactual scenario that captures the current approach of existing capacity expansion (CE) analyses[36,37], which ignore zoning regulations. Across all scenarios, we implement an 80% CO2 emission reduction target by 2040, aligning with state[38] and federal[39] policies as well as broader decarbonization goals[40] **(see Table 4 and Scenario Framework in Methods)**.

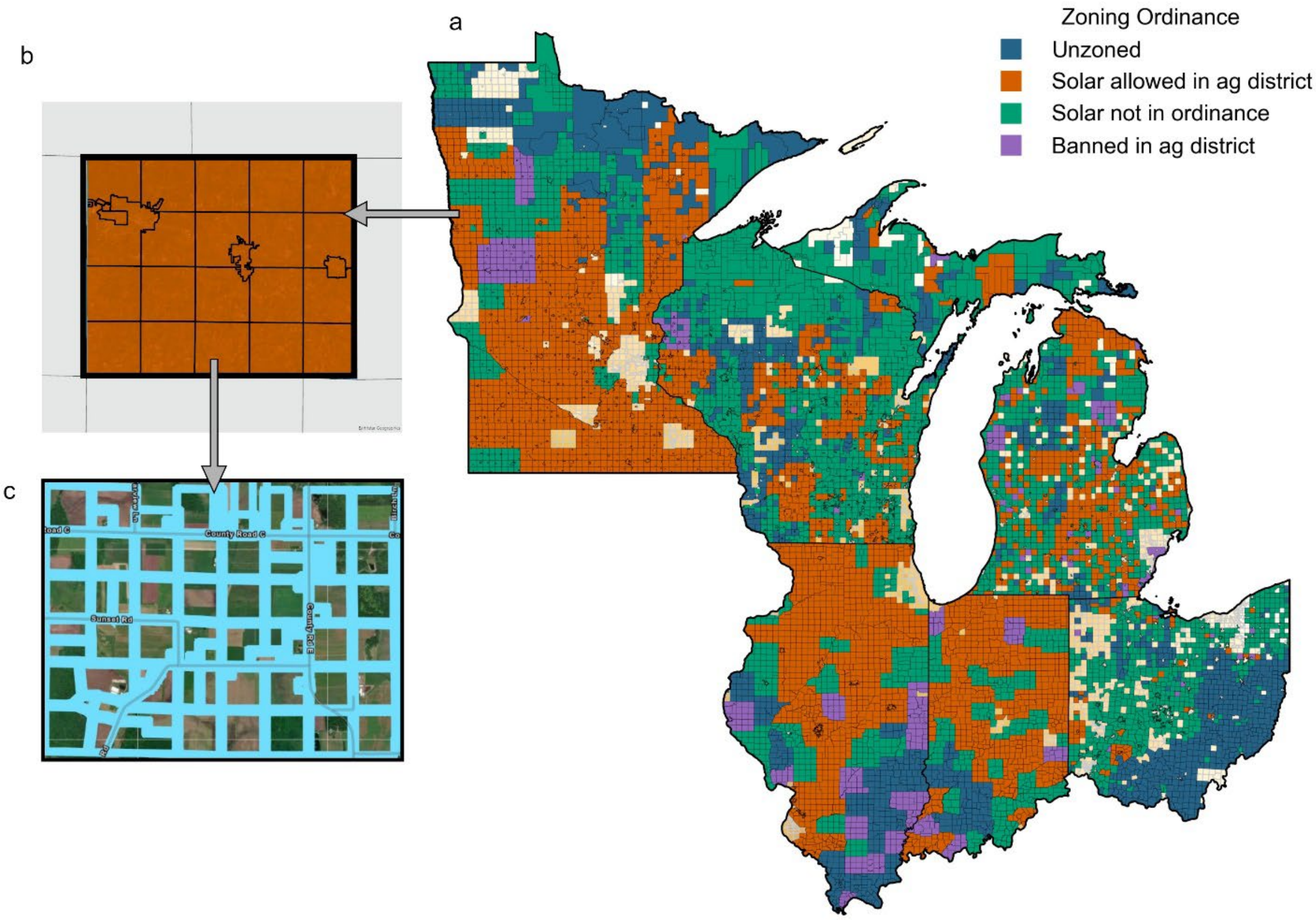

**Fig. 1.** a. Map showing regulations pertaining to utility-scale solar energy systems within subdivisions in the study area. b. Zoomed-in map showing a county-level zoning ordinance applied across county subdivisions. c. Zoomed-in map showing the implementation of zoning ordinances on land parcels within subdivisions where solar is allowed.

## Results

### Zoning ordinances in the Great Lakes Region

We collected 2,474 unique ordinances across the six Great Lakes states, covering both county and township/village-level jurisdictions, which we used to characterize the status of 9,997 subdivisions. The difference between the subdivisions (9,997) and the unique jurisdiction (2,474) arises because, in many cases, a single county ordinance governs multiple unincorporated subdivisions. For example, our data includes 399 county-level ordinances, mostly from Indiana and Illinois. In some cases, individual subdivisions also have their own ordinance, as is the case with 2,075 townships in Michigan, Minnesota, Ohio, and Wisconsin. **Table 1** summarizes the distribution of zoning authorities for all subdivisions in each state, along with the number of unique ordinances collected. The data reveal a diversity in zoning authority arrangements, for example, in Illinois, 65.8% of subdivisions are zoned at the county level, whereas in Michigan, the majority (75.2%) are governed by township-level ordinances. The database also captures how zoning ordinances address utility-scale solar energy systems—distinguishing among those that permit solar on agricultural lands, those that are silent on the issue, and those that explicitly ban solar development

**(Tables 2)**. Among the 7,679 rural subdivisions examined, 3,905 permit utility-scale solar on agricultural land, 3,262 ordinances are silent on the matter, and 512 subdivisions explicitly ban solar development in their agricultural districts. **Table 3** provides detailed insights into key zoning provisions such as setbacks and lot sizes. For instance, road setbacks range from 10 to 2,640 feet, with an average of 110 feet. Similarly, setbacks from non-participating property lines range from 5 to 500 feet, averaging 76 feet.

**Table 1.** Summary of subdivisions studied, zoning authority types, and individual ordinances collected. Note: The difference between the total number of subdivisions studied and the number of individual zoning ordinances collected arises because, in many cases, a single county ordinance governs multiple unincorporated subdivisions. "No Data" refers to zoning jurisdictions where researchers were unable to definitively determine which level of government had zoning authority.

| State | Total number of Subdivisions studied | Subdivision zoned by county | Subdivisions zoned by township/ /village | Unzoned subdivisions | No Data | Individual ordinances collected |
|---|---|---|---|---|---|---|
| Michigan | 1,518 | 216 (14.2%) | 1,142 (75.2%) | 130 (8.6%) | 30 (2.0%) | 1,281 |
| Ohio | 1,381 | 163 (11.8%) | 637 (46.1%) | 515 (37.3%) | 66 (4.8%) | 574 |
| Indiana | 1,010 | 683 (67.6%) | 228 (22.6%) | 90 (8.9%) | 9 (0.9%) | 82 |
| Illinois | 1,707 | 1,124 (65.8%) | 273 (16.0%) | 310 (18.2%) | 0 (0.0%) | 78 |
| Wisconsin | 1,760 | 1,286 (73.1%) | 262 (14.9%) | 205 (11.6%) | 7 (0.4%) | 288 |
| Minnesota | 2,621 | 2,396 (91.4%) | 95 (3.6%) | 104 (4%) | 26 (1.0%) | 171 |
| **Region** | **9,997** | **5,868** | **2,637** | **1,354** | **138** | **2,474** |

**Table 2.** Summary of subdivisions regulating large-scale solar on rural agricultural (ag) lands in the study area.

| State | Subdivisions that allow utility-scale solar on agricultural land/district | Subdivisions that do not mention utility-scale solar in ordinance | Subdivisions that ban utility-scale solar within the jurisdiction or in agricultural district |
|---|---|---|---|
| Michigan | 391 | 659 | 2/71 |
| Ohio | 56 | 549 | 15/24 |
| Indiana | 578 | 264 | 0/58 |
| Illinois | 784 | 310 | 37/215 |
| Wisconsin | 470 | 937 | 0/32 |
| Minnesota | 1626 | 543 | 1/112 |
| **Region** | **3,905** | **3,262** | **55/512** |

**Table 3.** Statistics of provisions/regulations in the 2,474 zoning ordinances surveyed. Number indicates number of ordinances collected (of the 2,474 in Table 2) that include a given provision.

| | Number | Average | Mode | Range |
|---|---|---|---|---|
| Setback: Road (feet) | 229 | 110 | 100 | 10-2640 |
| Setback: Participating property line (feet) | 280 | 53 | 50 | 0-500 |
| Setback: Non-participating property line (feet) | 328 | 76 | 50 | 5-500 |
| Setback: Participating residence (feet) | 151 | 126 | 0 | 0-1320 |
| Setback: Non-participating residence (feet) | 218 | 265 | 200 | 40-2500 |
| Minimum lot size (acres) | 186 | 15 | 5 | 0-500 |
| Maximum lot size (acres) | 9 | 70 | 40 | 2.5-250 |
| Maximum lot area coverage (%) | 101 | 72% | 100% | 5-100% |

### Impact of zoning ordinances on utility scale PV technical potential

We first examine how each of our zoning scenarios **(Table 4)** affect utility-scale PV supply curves for the entire study region **(Fig. 2a)** and states **(Fig. 2b).** In the current zoning scenario, ordinances silent on solar—which in these "permissively" constructed zoning ordinances are de facto bans[20]—pose the largest reduction in available capacity for utility-scale PV deployment **(Fig. 2, SI Figs S10 and S11)**, reducing available capacity by 31% (940 GW) across our region and by up to 59% (177 GW) at the state level. Outright bans—either by explicitly disallowing solar installations or by not permitting solar in the predominant agricultural zoning district—contribute another 7% reduction across the region and up to 14% reduction at the state level. Non-participating property line and road setbacks are the next two largest sources of capacity reductions for utility-scale PV deployment, each reducing available capacity by up to 6% across our region **(Fig. S10)**. When accounting for all setback and lot size requirements, the available capacity is reduced by 16%, which is substantially smaller than the reductions caused by ordinances silent on solar installations alone. This contrasts with research by Lopez et al. who identify setbacks as the major contribution to capacity reduction[26]. Altogether, zoning regulations, including silent ordinances, contribute to a 54% (or 1.6 TW) decrease in total available utility-scale PV capacity across our study region **(Fig. 2a)**. At the state level, available capacities decrease from 45% (or by 323 GW) in Minnesota to 76% (or by 216 GW) in Ohio (**Fig. 2b**). Even with these reductions in the current zoning scenario, significant solar capacity investment potential (over 79 GW) remains in each state, and existing zoning ordinances modestly increase the cost of supply across the entire study area, states and load regions. The cost of deploying 5 GW of utility-scale PV in Minnesota **(Fig. 2b)**, for instance, increases by 1% (or by $0.8/MWh) from the ignoring zoning to current zoning scenario.

In the expanded zoning scenario, which effectively eliminates capacity reductions caused by silent ordinances **(Table 4)**, additional capacity reductions still occur due to new bans, setbacks, and lot size restrictions **(Fig. 2)**. Nevertheless, the expanded zoning scenario offers 663 GW more potential capacity **(Fig. 2a)** than the current zoning scenario.

**Table 4.** Scenario Framework. Each scenario is run with an 80% $CO_2$ emission reduction target by 2040.

| Scenario | Description |
|---|---|
| Ignoring zoning | Assumes all subdivisions are available for solar development without applying zoning ordinances; serves as a counterfactual, reflecting the typical approach used by existing capacity expansion (CE) analyses[36,37]. |
| Current zoning | Applies existing zoning ordinances to subdivisions, including both permissive and outright bans; permissive areas are subject to setbacks and lot size restrictions as defined by the ordinances; unzoned areas remain developable[20] with average setbacks and lot sizes applied (**Table 3**); excludes solar in areas that are silent. |
| Expanded zoning | Builds on the current zoning scenario by replacing silent ordinances with either permissive or outright bans; in permissive areas, it includes details such as setbacks and lotsize restrictions. |

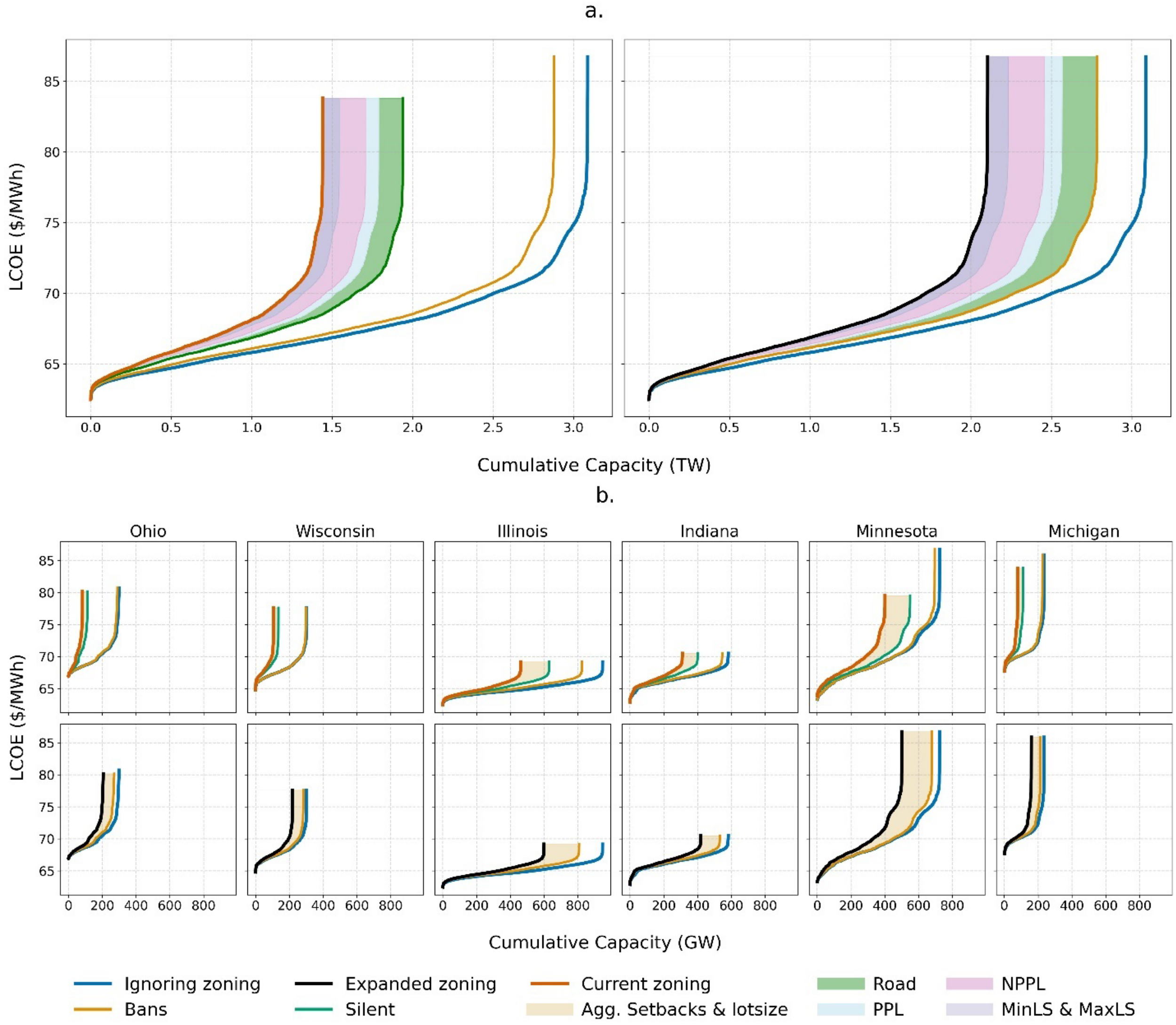

**Fig. 2.** Change in supply curves for the entire study area (a) and across states (b) from the ignoring zoning to current zoning and expanded zoning scenarios. We illustrate changes in the supply curves by considering only outright bans (light orange line) or both outright and de facto bans (green line). Additionally, we illustrate the additive effects of road setbacks, participating property line (PPL) setbacks, non-participating property line (NPPL) setbacks, and minimum and maximum lot size requirements (MinLS and MaxLS, respectively), resulting in cumulative capacity under the current zoning scenario (deep orange line) and the expanded zoning scenario (black line). These impacts are aggregated across states in (b) but separated in the SI **(Figure S8).**

### Impact of zoning ordinances on utility-scale PV investments and costs

We analyze the impacts of zoning ordinances on rural utility-scale PV investments by running our CE model for the “ignoring zoning”, “current zoning” and “expanded zoning” scenarios, while accounting for the 80% carbon emission reduction target by 2040 (**Table 4**). Across our three scenarios, most investments occur in wind and solar capacity, which expand by 112 GW and 35-43 GW, respectively (**Fig. 3; and SI Fig. S12**), by 2040. Accounting for solar zoning ordinances shifts investment between generation types. In the current zoning scenario, installed utility-scale capacity decreases by 18% (8 GW) compared to the ignoring zoning scenario **(Fig. 3)**. This reduction stems mainly from ordinances silent on solar effectively blocking solar development in key competitive sites that benefit from lower levelized costs of electricity (LCOE) due to their higher capacity factors. Investment primarily shifts from solar to natural gas combined cycle (NGCC) and energy storage, which increase by 10% (3 GW) and 8% (900 MW), respectively.

In contrast, the expanded zoning scenario leads to a smaller reduction of 3% (1.5 GW) in installed utility-scale solar capacity compared to the ignoring zoning scenario **(Fig. 3)**. This smaller decrease is solely attributable to explicit bans and reduced land availability due to setbacks and lot-size restrictions in competitive solar sites, rather than also due to silent ordinances. Given this modest reduction in solar capacity, only marginal increases in investments in NGCC and energy storage are observed.

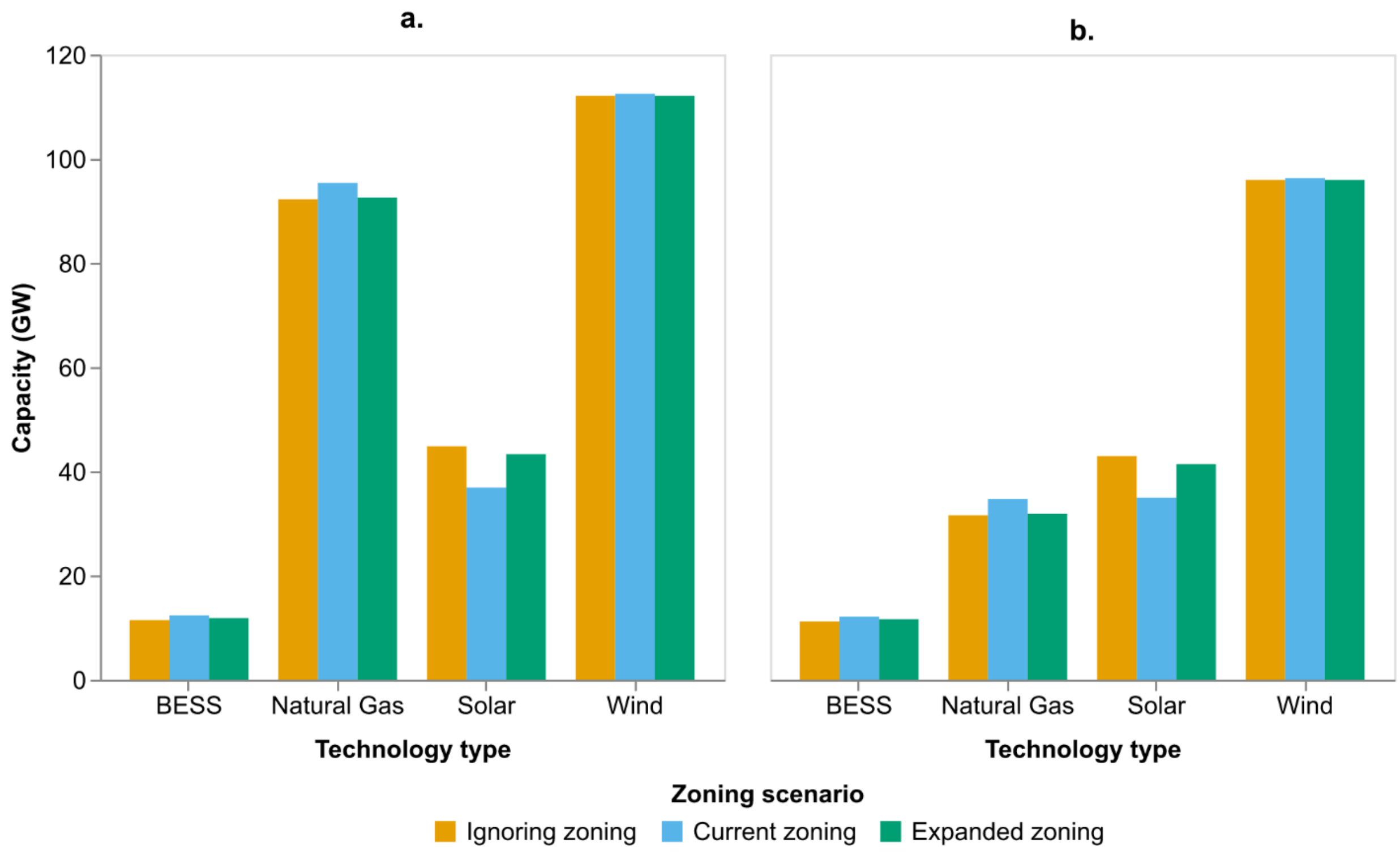

**Fig. 3.** Capacity by technology through 2040 under zoning scenarios to meet an 80% carbon emission reduction target. a. Total capacity and b. New investments for battery energy storage (BESS), natural gas, solar, and wind power.

**Fig. 4** illustrates the geographic distribution of new utility-scale PV deployments in the ignoring zoning scenario **(4a)** and shifts from the ignoring zoning to current zoning **(4b)** and expanded zoning **(4c)** scenarios. Among other factors, utility-scale PV investments occur in high solar resources areas within each state near transmission lines (e.g., in southern Michigan and southwest Illinois and Indiana) **(see SI Methods, Fig. S7)**. As zoning ordinances are applied in the current and expanded zoning scenarios, the available land for PV investments diminishes, restricting capacity investment in the most preferrable investment locations, driving investment to new sites. The current zoning scenario, which contains silent ordinances, eliminates investment in 80% (161) of locations that had investments under the ignoring zoning scenario **(4b)**. In comparison, the expanded zoning scenario only eliminates investments on 41% (83) of previously viable locations, as silent ordinances are replaced by a mix of permissive and restrictive solar ordinances. Average investment per site shrinks by 20% for the current and 36% for expanded zoning scenarios, driven by zoning-constrained area available for development. The expanded zoning scenario, by expanding investment sites for PV capacity relative to the current zoning scenario, results in less of a shift in where new investments occur.

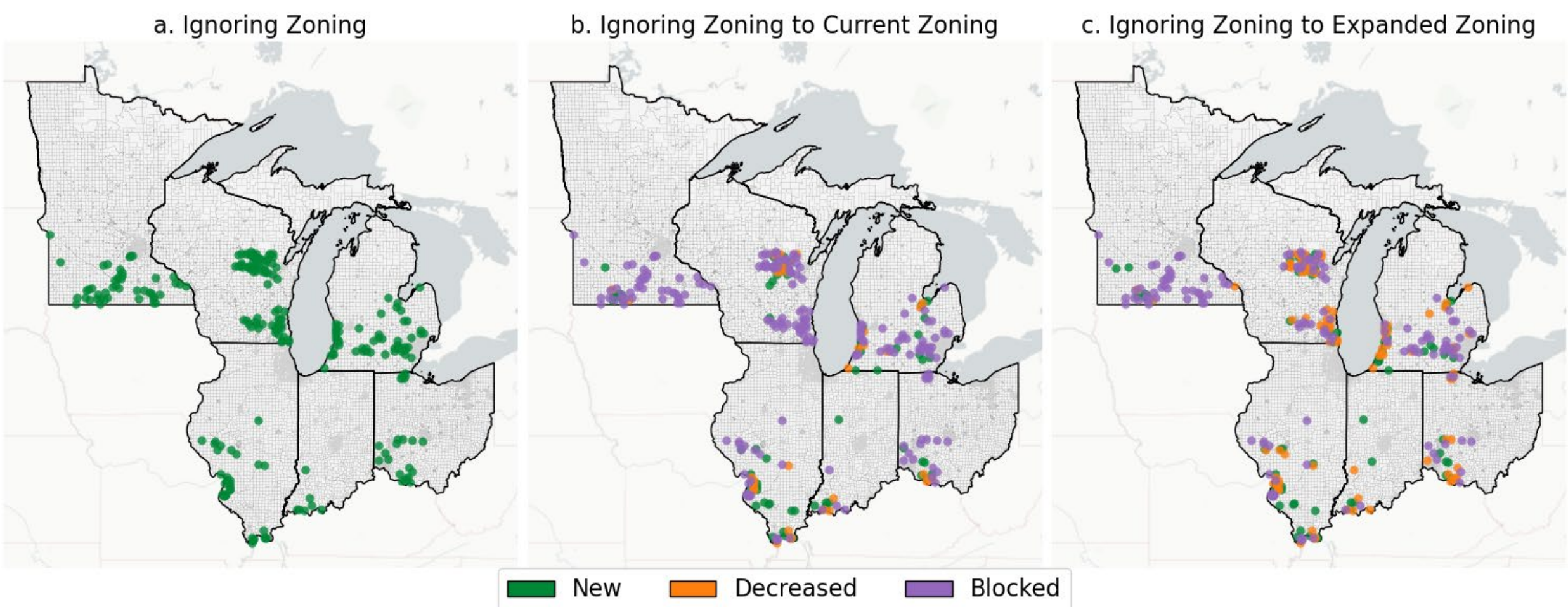

**Fig. 4.** Geographic distribution of utility-scale PV deployments under the ignoring zoning scenario (a), along with changes in PV deployments from the ignoring zoning to current zoning scenario (c) and from the ignoring zoning to expanded zoning scenario (d). New investments are in green, investments with capacity reduction in orange (due to zoning regulations), and investments that are blocked (largely due to silence or explicit bans) in purple.

Site-level shifts in solar PV investments accumulate to more solar PV investments in states with less zoning restrictions and more attractive (i.e., high solar resource and low cost) sites **(Fig. 5, Fig.2)**. Under the current zoning scenario, solar capacity investments decrease by 3-16% in Illinois and Ohio and 39-42% in Michigan and Wisconsin, compared to the ignoring zoning scenario **(Fig. 5)**. This reduction is due to the combination of higher deployment costs in these states (**Fig. 2**) and the prevalence of silent ordinances blocking solar development **(Fig. 1)**. Specifically, Michigan, Wisconsin and Ohio are observed to have more than 50% of total land area prohibited from development under the current-zoning ordinance, placing them among the highest in terms of restrictions **(Fig.1, see also SI Fig. S11)**. Conversely, Indiana and Minnesota, with the lowest deployment costs and fewer bans, experience significant increases in solar deployment—11% and 120%, respectively—across zoning scenarios. Similarly, in the expanded zoning scenario, capacity investments align with state supply curves **(Fig. 2)**. Here, investments largely change locations within each state, rather than shifting between states.

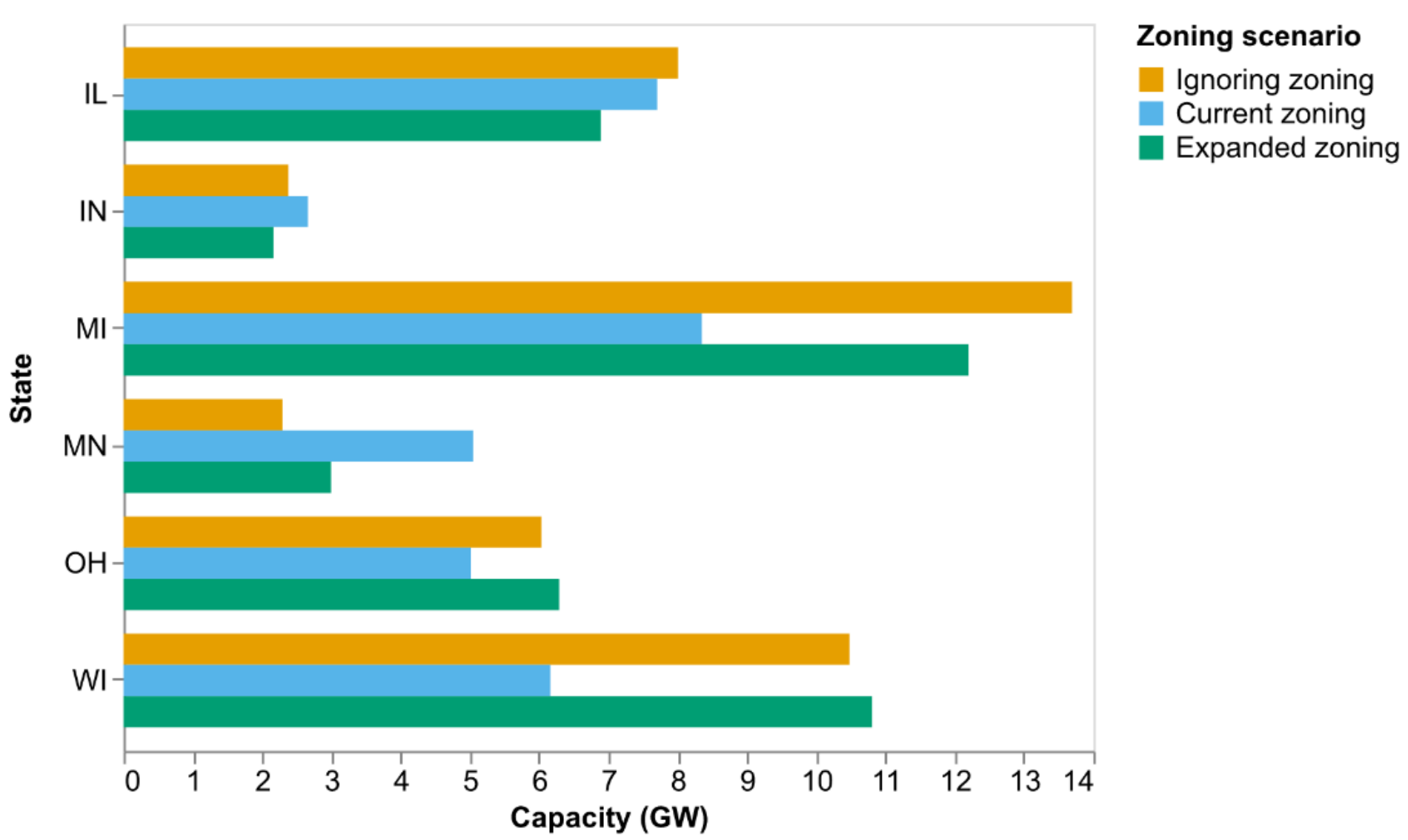

**Fig. 5.** Change in capacity across states with the implementation of zoning scenarios.

Relative to the ignoring zoning scenario, enforcing of zoning ordinances reduces utility-scale PV investment by 18% ($4.8 billion) in the current zoning scenario, and by 3% ($900 million) in the expanded zoning scenario. Investment costs decline linearly with reduced capacity investments. As investment shifts from solar to natural gas generators and energy storage due to zoning ordinances **(see Fig. 3),** total fixed and variable investment costs increase by 1% ($1 billion). In contrast, changing from ignoring zoning to expanded zoning has minimal effect on the total technology investment cost.

**Table 5.** Total cost of investment in utility-scale PV across planning years for zoning scenarios and utility-scale PV deployment scenarios

| Cost category | Cost per zoning scenario (billion USD) | | | Percentage change | |
|---|---|---|---|---|---|
| | Ignoring zoning | Current zoning | Expanded zoning | Ignoring zoning to Current zoning | Ignoring zoning to Expanded zoning |
| PV investment cost | 26 | 22 | 25 | 18% | 3% |
| Total investment cost (Fixed + variable cost) | 142 | 143 | 142 | 1% | 0 |

## Discussion

This paper quantifies the impact of zoning ordinances on power sector decarbonization in the Great Lakes region of Illinois, Indiana, Michigan, Minnesota, Ohio, and Wisconsin, focusing on utility-scale PV deployment. We integrate a database of 2,474 unique rural zoning ordinances for utility-scale solar, spanning 9,997 subdivisions (more than 90% of county subdivisions in each state), with a power system planning (or capacity expansion) model driven by an 80% carbon emission reduction by 2040 **(ref Scenario**

**Framework)**. Across our study region, accounting for existing zoning ordinances significantly affects regional and state-level solar potential and deployment, largely driven by "silent" zoning ordinances that implicitly block solar development by failing to address its land-use[20]. Specifically, our study shows a 31% reduction in available utility-scale PV capacity (940 GW) across the region, with reductions reaching up to 59% (177 GW) at the state level. This reduction results in an 18% decline in utility-scale PV investment by capacity (8 GW) and cost ($4.8 billion) **(Figs. 2-4 and Table 4).** At the state level, there is a notable shift in development patterns. States with a high prevalence of zoning ordinances that are silent on solar, such as Michigan and Wisconsin, see utility-scale PV investments decrease by up to 42%. In contrast, states with ordinances that explicitly ban or allow solar development, such as Indiana and Minnesota, attract more investment **(Figs. 1, 4 and 5)**. Our study indicates that while zoning ordinances impose certain restrictions on solar development—such as setbacks and minimum or maximum lot size requirements, as outlined in **Table 4**—these constraints are far less restrictive than silent ordinances. Even though zoning ordinances may result in some areas being taken off the map entirely, this is an exception rather than the rule.

Our results have significant implications for state and local policymakers. Increasingly, attention is being paid to restrictive local zoning ordinances as a bottleneck to solar deployment[41]. Analysis finds that there is some truth to that:  there are some explicit bans. But much more common and detrimental to solar deployment are zoning ordinances that are silent on solar development, which create a regulatory gray area, leading to unintentional barriers to utility-scale PV deployment. Several case studies[42] have alluded to undefined solar in zoning causing project delays and increasing costs, as developers navigate uncertain regulatory landscapes. One way that many states are addressing this issue is by removing local zoning authority over large projects. Indeed, over the course of our research two of our six states adopted regulations that significantly constrained zoning authority for local governments. Such an approach, though, is not without drawbacks, which may include electoral backlash[43] and a rise in urban-rural conflict[44]. An alternative approach may be to instead require localities to proactively update their zoning ordinances in consultation with stakeholders, to eliminate the unintended obstacle posed by silence.

While zoning decisions are made at the local level, our study indicates they significantly influence the distribution and scale of utility-scale PV deployment across larger regions. Indeed, the move to shift siting authority to the state level recognizes that local regulations may impede statewide decarbonization goals. Our study highlights, though, how local and state zoning decisions can have multifaceted impacts on wider energy objectives throughout the region. While we do not delve into the specific impact of solar capacity shifts between states, previous research by Moore et al. suggests that it may result in competition for valuable agricultural lands and increased land acquisition costs [45]. Additionally, zoning-induced shifts in PV deployment would shift economic activity at the state and local level, with potentially long-term consequences for renewable investment [46]. In the short-term, shifts in PV deployment would also move utility-scale PV-related jobs in construction, installation, and operation [47,48]. Increased investment and jobs in communities hosting utility-scale PV installations can increase local taxes and community revenues. However, the uneven distribution of these benefits can exacerbate regional disparities in energy services and community development. Policymakers must understand the effect of zoning ordinances to formulate achievable and equitable decarbonization plans. Moreover, the need for compensatory investments in natural gas due to reduced solar investments underscores the importance of supportive zoning in sustaining the progress towards a low-carbon energy system. Zoning reforms that provide clarity on solar development can ensure that solar projects are developed efficiently and at scale, preventing lock-in of fossil fuel infrastructure.

Our research extends the work of Lopez et al. [26], who surveyed a smaller range of zoning jurisdictions and provisions (such as setbacks or lot size restrictions), then extrapolated those provisions to other regions or counties where local zoning ordinances were not specifically surveyed. In contrast, we use an extensive zoning database to capture zoning ordinances in jurisdictions that are silent on solar; implement solar bans and permissions; use actual setback distances instead of representative ones; and consider minimum and maximum lot size restrictions [26]. Additionally, while Lopez et al. [26] focus primarily on the impact of setbacks, we find that silence in states with permissive zoning alone could lead to a capacity reduction of up to 31% across our region and 59% at the state level (Fig. 1). This suggests a substantial decrease in available utility-scale PV capacity across various states than the 18% capacity reduction reported by Lopez et al [26]. Thus, our study reveals a more pronounced impact of zoning ordinances than previously understood.

Our study has several limitations that future research could examine. First, we consider utility-scale PV deployment solely on agricultural lands. This approach was informed by initial assessments showing that that 60% of existing utility-scale solar facilities in our study area have 80-100% of their site located on agricultural land **(see SI Fig. S5)**. However, for geographical scopes wider than our study region, barren and contaminated lands may offer a promising opportunity for deploying utility-scale PV to minimize land-use conflicts and environmental impacts [9,31]. Second, we do not assess how restrictive wind ordinances in these same geographies may shape renewables deployment. Our focus is primarily on solar, as it is more significantly impacted by zoning ordinances in agricultural or rural areas where land-use conflicts are common[49]. Future research beyond the rural agricultural scope should consider both solar and wind ordinances simultaneous in these geographies. Thirdly, we do not distinguish between various zoning approval mechanisms, such as by-right use, special exceptions, and conditional uses. We anticipate that these varying levels of permission will further increase investment costs and reduce capacity investments, as the level of scrutiny rises. Future research should address this gap by examining how different levels of zoning permissions affect utility-scale solar development to provide a more comprehensive understanding of the regulatory landscape and its impact on project deployment.

Our study focuses on zoning ordinances that directly impact land availability for solar. But there are additional regulations housed within existing zoning ordinances or outlined in Subdivision and Land Development Ordinances (SALDOs) or Special Land-use Permits that may not restrict land availability outright, but impose requirements that affect the design, construction, permitting process, and costs of solar projects. Such requirements may include provisions for stormwater management, fencing, vegetative screening, and decommissioning plans. Although we expect the costs associated with these regulations to be smaller and less impactful than those that limit land availability or prohibit solar development altogether, it is important for future research to investigate the cumulative and long-term impacts of these factors in more detail. Finally, it is imperative for future research to extend our investigation beyond our study region. Local governments in many other U.S. states where solar PV is rapidly growing, including North Carolina, Texas, California, and Arizona, also have local government zoning control for or all large-scale solar projects[50]. Enlisting a coalition of universities and practitioners to supplement NREL's existing database will allow for national assessments of ordinances and their impacts on decarbonizing the United States.

## Materials and Methods

Our analytical framework is summarized in Fig. 6 and described briefly below. Please refer to SI Methods for a detailed description.

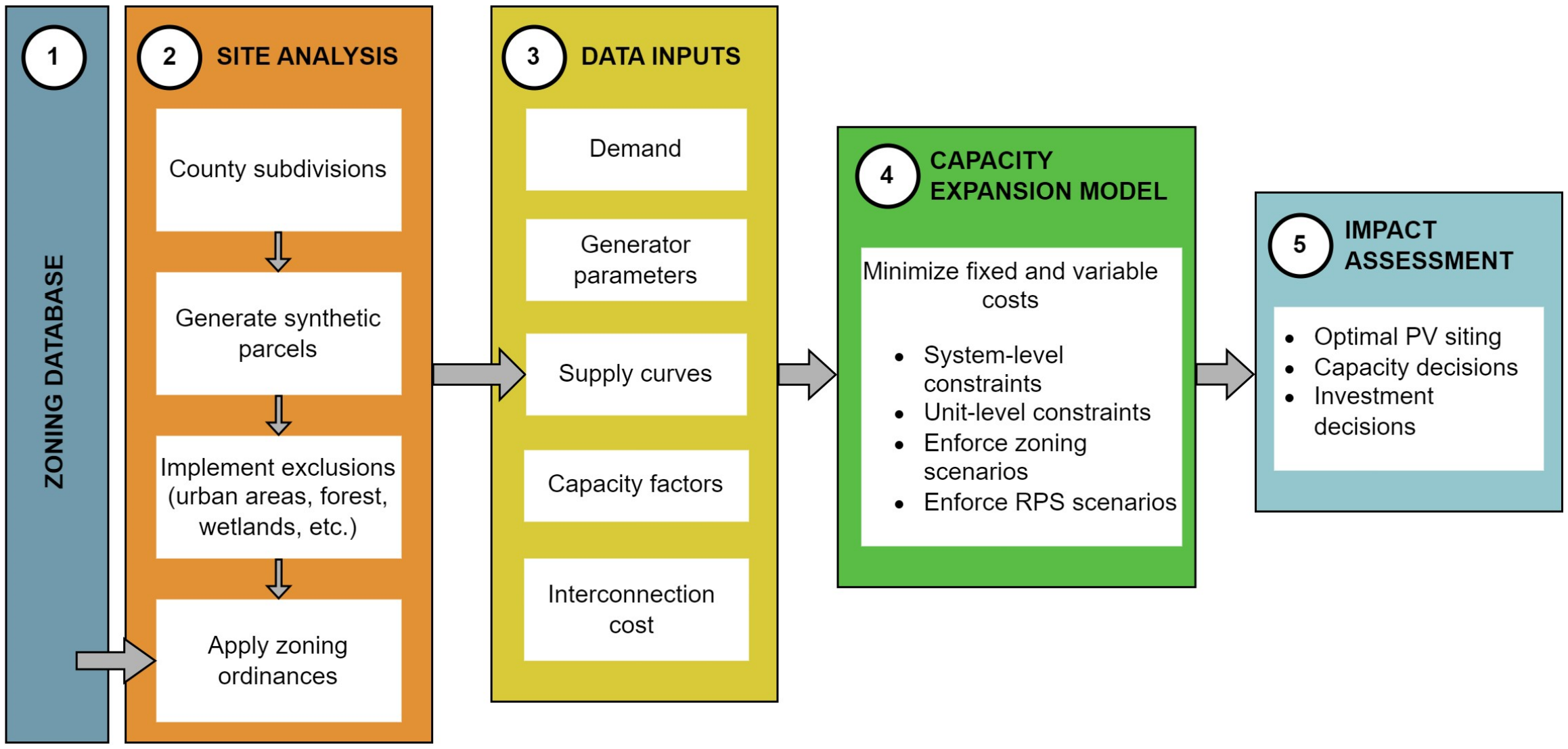

**Fig. 6.** Analytical framework used in this study.

### Quantifying land area available for utility-scale PV deployment given zoning ordinances.

To estimate land area available for utility-scale PV deployment, we apply zoning ordinances and land use exclusions to individual land parcels. The foundation of our study was the development of a database of 2,474 solar zoning ordinances across our Great Lakes study region of Illinois, Indiana, Michigan, Minnesota, Ohio, and Wisconsin. The data includes of 399 county-level ordinances, mainly in Indiana and Illinois and 2075 subdivision-level ordinance in Michigan, Minnesota, Ohio, and Wisconsin. This database is unique in its scale and level of detail. Zoning authority (categorized as either county, city, township, village, other or unzoned) was established for each jurisdiction through public records, communication with local municipalities, county master plans, and expert input. For each authority, ordinances were collected through online, e.g. through a municipality's website or Municode, or offline means, e.g. through direct procurement. Ordinances were then manually searched for mentions of solar energy. Any mention of solar energy was categorized as to whether it applied to solar as a principal and/or accessory use; principal use regulations were then classified by 17 characteristics, including various setbacks and lot size limits. The zoning ordinances were subsequently used to characterize the status of 9,997 county subdivisions. The difference between the 9,997 subdivisions and the 2,474 unique ordinances collected arises because, in several cases, a single county ordinance governs multiple townships. In some of these states, state-level authority for siting large utility-scale solar projects rests not with local governments but with state authorities. The threshold of what constitutes large varies from 50 MW in Ohio and Minnesota to 100 MW in Wisconsin [51]. In each of these states, local governments are given discretion to regulate smaller utility-scale solar projects, and in some states (notably Ohio), local governments are given some opportunity to influence the state-level process. Furthermore, solar siting policies in the region are in flux, e.g. solar siting authority in both Illinois and Michigan changed in the process of this research. In all cases, we chose to include and classify the ordinance for the largest principal-use solar under a jurisdiction's control. We also considered a jurisdiction to allow utility-scale solar if the zoning ordinance permitted it through any regulatory mechanism, including by-right, special exceptions and conditional uses. We did not differentiate between these categories due to the rarity of utility-scale solar being permitted as a by-right use and the extensive resources required to classify each ordinance accordingly. The database is publicly accessible through the University of Michigan's Graham Sustainability Institute's website [24].

We quantify land available for utility-scale PV at each county subdivision by applying land exclusions and zoning ordinances on individual land parcels. Due to the lack of comprehensive statewide parcel data, we leverage available parcel data from Wisconsin and Indiana to generate synthetic parcels using geospatial analysis in ArcGIS Pro and nearest neighbor analysis [52,53] **(SI Section 1.1)**. We validate our result by comparing the size and distribution of our synthetic parcels to the size and distribution of actual parcels which we obtain from the Lawrence Berkeley National Laboratory (LBNL) (SI Fig. S2). To focus solely on utility-scale PV potential in rural areas and agricultural lands, we exclude urban areas, urban areas buffers, restricted lands, and non-agricultural lands using data on urban land data from the US Census Bureau's, Cropland Data Layer from the United States Department of Agriculture, and Prime Farmland data from the USDA Natural Resources Conservation Service (NRCS) [54–56] **(SI Section 1.1).** This method aligns with current trends in utility-scale PV development, where agricultural lands are frequently prioritized due to their flat topography and accessibility[34] **(SI Figure S5)**. Additionally, we also exclude sites with slope unsuitable for wind or solar development, that is, 10 degrees for solar and 19 degrees for wind, with thresholds based on research by Leslie et al. and Wu et al[9,10].

We map subdivisions enclosing our parcels to the database of renewable energy zoning ordinances to assess land availability for utility-scale PV under various zoning regulations. We categorize subdivisions as either "zoned" (with zoning ordinances) or "unzoned" (without zoning ordinances). Among the "zoned" subdivisions, we further differentiate between those that regulate principal-use solar energy systems and those that do not (i.e. "silent"). For "zoned" subdivisions that permit in their dominant agricultural district, we enforce road setbacks, participating property line setbacks (PPL), non-participating property line setbacks, and minimum and maximum lot size restrictions on individual parcels **(Fig. 7)**. In the current zoning scenario, we treat "silent" zoned ordinances as de facto bans since zoning in each of these states is constructed permissively—that is, land use is assumed to be not allowed unless it is expressly mentioned in the zoning ordinance. In the expanded zoning scenario, we assume these "silent" jurisdictions will amend their ordinances, mimicking the permissions, setbacks and lot-size regulations used in the "zoned but not silent" jurisdictions **(Table 3)**. For "unzoned" subdivisions, we apply average setbacks and lot-size regulations in present in the zoned subdivisions **(Table 4)** to address the absence of zoning ordinances. Subsequently, we aggregate all available land by subdivision.

### Assessing solar resource, interconnection cost and distance by subdivision

We estimate the mean annual hourly capacity factors for each subdivision using hourly solar insolation, air temperatures, and other relevant meteorological variables from the National Solar Radiation Data Base (NSRDB) [57], coupled with simulations from the System Advisor Model (SAM) [58] **(SI section 1.3.1)**. Furthermore, we estimate spatially explicit costs for connecting utility-scale PV facilities in each subdivision to the bulk transmission grid by measuring the shortest straight-line distance. Our approach accounts for both electrical components and right-of-way costs and incorporates transmission line data from the Homeland Infrastructure Foundation Level Database **(SI section 1.3.2)**. We combine these interconnection costs with comprehensive capital and operational expenditure costs, as detailed in NREL's 2021 ATB [59] utilize the "Market" case within the moderate scenario, to estimate the levelized cost of energy (LCOE) at each subdivision. The assumptions underlying the ATB and LCOE estimates are detailed in **the SI Table S6.** Capacity factors and costs characterizing solar development at each subdivision were used to construct the supply curves, which were subsequently integrated into our capacity expansion model.

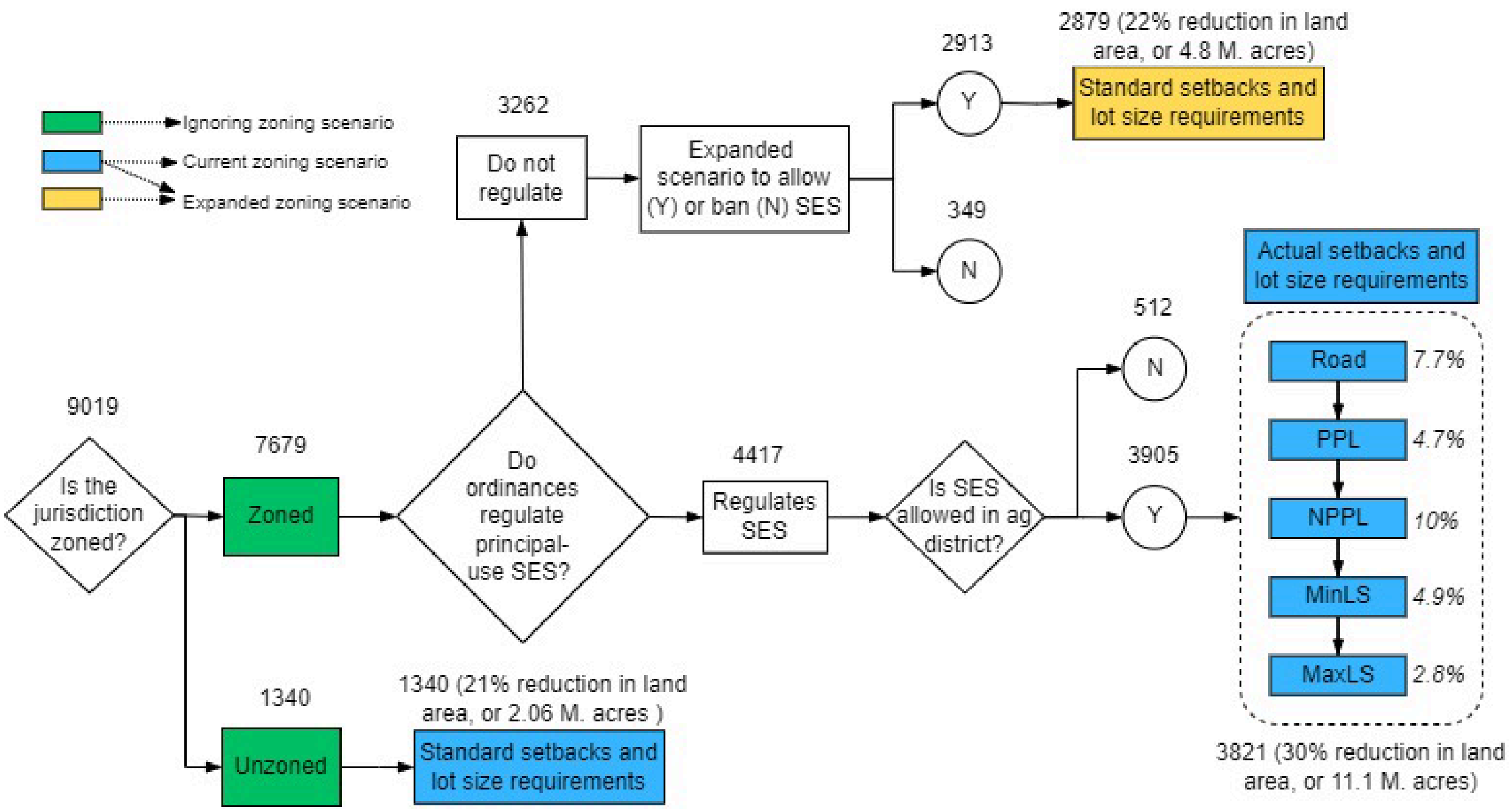

**Fig. 7.** Mapping of zoning regulations for principal-use solar energy systems, or utility-scale PV systems. Subdivisions that allow within their dominant agricultural district are labeled as "Y" and those that prohibit in their agricultural district are labeled as "N". Setbacks from roads are labelled as "Road", those from participating property lines as "PPL", and from non-participating property lines as "NPPL". Minimum and maximum lot size requirements are labelled as "MinLS" and "MaxLS", respectively. The count tracks the number of subdivisions where ordinances are implemented, while the percentages indicate the land area reduction from regulations that limit lot size. Boxes are shaded by the scenario in which they are included, as detailed in the Scenario Framework section (green: ignoring zoning, blue: current zoning, and yellow: expanded zoning).

### Capacity expansion model

To determine the optimal utility-scale PV capacity at each subdivision, we use a long-term planning, or capacity expansion (CE), model. The CE model optimizes generator and transmission investments and hourly operations while minimizing total annualized system costs. Total system costs comprise the sum of annualized fixed investment and variable operating cost. Fixed costs account for the capital and fixed operations and maintenance (O&M) costs of new transmission lines, electricity generators, and storage. Variable costs account for fuel and variable O&M costs associated with operations of new and existing units. Fixed and variable cost projections are obtained from the US National Renewable Laboratories (NREL)'s 2020 Electric Annual Technology Baseline [59] **(SI Section 3, Appendix 2)**. We account for the provisions of the Inflation Reduction Act of 2022, by incorporating a 30% Investment Tax Credit (ITC) on the capital costs of solar **(Table S6)**. We assume that all eligibility requirements for the 30% ITC, including prevailing wage, project size and apprenticeship provisions, are met.

The CE model is run from 2020 to 2040 in 5-year time steps. We divide our region into 24 subregions, between which limited transmission capacity exists [37]. Within each time step, the model chooses investments in solar PV, wind, coal steam with carbon capture and sequestration (CCS), natural gas combined cycle (NGCC), NGCC with CCS, nuclear power plants, battery storage and transmission lines. Solar PV and wind investments occur at the county subdivision, while all other investments occur at the subregional level. For solar, we establish the maximum PV capacity threshold based on the area of agricultural lands within each rural subdivision after implementing zoning ordinances. Similarly, wind is deployed exclusively on agricultural lands. We convert land area to potential capacity for wind and solar using wind and solar power densities of 0.5 $Wm^{-2}$ and 5.4 $Wm^{-2}$, respectively [60]. We obtain future capital and operational costs from NREL's Annual Technology Baseline[59]; interregional transmission capacity from NREL's Regional Energy Deployment System (ReEDS) Model [37]; and future electricity demand profiles by subregion for a moderate electrification future [61]. Wind, solar, and demand timeseries correspond to 2012 meteorology, capturing co-variability in these meteorologically-dependent variables. A comprehensive list of our input datasets and assumptions is detailed in **the SI Table S6.** Our initial generation fleet is detailed in **Table S7.**

The CE model enforces several system-level constraints, including balancing regional electricity and demand and meeting reserve requirements on an hourly basis **(SI Section 3, Appendix 2)**. The CE model also requires all zoning scenarios to meet an 80% $CO_2$ emissions constraint by 2040 (ref Scenario Framework below). The model also incorporates unit-level constraints, e.g., upper limits on the generation capacity of power plants **(SI Section 3, Appendix 3)**. We model power flows between load regions using the simple transport model, which optimizes flows between pairs of regions under a maximum interregional transmission capacity [62]. For computational tractability, each modeled year is subdivided into two representative days per season and the day with the peak annual demand [63] (SI). The CE model is implemented in GAMS and solved using CPLEX. To reduce the computation time, we select the top 20% wind and solar sites in our study region based on their levelized cost of transmission.

Transmission lines that connect load regions within each state are combined into a single bidirectional line. This line has a capacity equivalent to the total capacity of all the individual lines connecting the two load regions. For simplicity, we assume that the distance (in miles) of each transmission line is equal to the distance between the centroids of the load regions it connects. Our capacity expansion (CE) model allows for the expansion of transmission line capacity to reduce congestion and support new generation facilities. This expansion is optimized by minimizing the cost per megawatt (MW) of additional transmission capacity, calculated by multiplying the per MW-mile cost of the line by its distance. Within load regions our CE model also optimizes the spatially-explicit costs of connecting new facilities to the bulk transmission grid **(SI Section 1.3.2)**.

### Scenario Framework

To quantify the impact of zoning ordinances on utility scale PV deployment, we assess three distinct zoning scenarios **(Fig. 7)**. The "ignoring zoning" scenario uses parcels from all zoned and unzoned subdivisions without implementing zoning ordinances. It serves as a counterfactual, reflecting the typical approach used by existing capacity expansion (CE) models[36,37] that often use broad land-use exclusions (e.g., protected areas, water bodies) but stop short of incorporating any zoning regulations when planning new utility-scale solar projects. The “current zoning scenario” applies existing zoning ordinances to parcels on a subdivision basis. For unzoned subdivisions, since there are no zoning ordinances, this scenario applies average setbacks and lot size requirements for utility-scale PV in our database **(refer to Table 4)**. In the zoned subdivisions, this scenario applies road setback, property line setback, and minimum and maximum lot size requirements expressly outlined in their zoning ordinances. It treats zoned subdivisions which are silent on principal solar as a ban[20]. The “expanded zoning scenario” reflects the ongoing adoption of utility-scale PV ordinances by communities. It assumes that areas with existing zoning ordinances that are currently silent on solar (omit solar) energy will update their regulations to explicitly address solar, either by allowing or banning it. We model this scenario by analyzing the distribution of subdivisions that explicitly allow or ban solar development, then apply this distribution to the subdivisions with ordinances that are silent on solar, treating them as if they had updated their zoning to either permit or ban solar. These updated subdivisions are added to the “current zoning” scenario to form the “expanded zoning” scenario **(Fig. 7, SI Section 1.1)**. For each zoning scenario, we set our capacity expansion model to meet a CO2 emission reduction target of 80% by 2040, with a linear interpolation of CO2 cap from 2025 to 2040. We set this target to aligning with state[38] and federal[39] policies, as well as broader decarbonization goals[40] to achieve net-zero emissions by 2050 [40].

**Acknowledgments**

This development of the zoning database at the University of Michigan was led by Sarah Mills and Madeleine Krol, with contributions from research assistants Leah Adelman, Thomas Banks, Geoffrey Batterbee, Sophie Farr, Rikki Goldman, Olivia Hintz, Samantha King, Zoe Knaus, Eric Li, Andrew Light, Zona Martin, Moksha Menghaney, Jon Newman, Abby Randall, Hayley Sakwa, Kaitlyn Sledge, Emma Uebelhor, and Josephine Weissburg.

This material is based upon work supported by the U.S. Department of Energy’s Office of Energy Efficiency and Renewable Energy (EERE) under the Solar Energy Technologies Office Award Number DE-EE0009361. The views expressed herein do not necessarily represent the views of the U.S. Department of Energy or the United States Government.